\documentclass[10pt,aps,prc,twocolumn,floatfix,showpacs]{revtex4-1} 
\usepackage{graphicx}
\usepackage{sidecap}
\usepackage{array}
\usepackage{subfigure}
\usepackage{color}
\usepackage{lineno}

\begin{document}%

\title{Observation of an energy-dependent difference in elliptic flow between particles and anti-particles in relativistic heavy ion collisions}

\author{
L.~Adamczyk$^{1}$,
J.~K.~Adkins$^{23}$,
G.~Agakishiev$^{21}$,
M.~M.~Aggarwal$^{34}$,
Z.~Ahammed$^{53}$,
I.~Alekseev$^{19}$,
J.~Alford$^{22}$,
C.~D.~Anson$^{31}$,
A.~Aparin$^{21}$,
D.~Arkhipkin$^{4}$
% ...
(STAR Collaboration)
}

\address{$^{1}$AGH University of Science and Technology, Cracow, Poland}
% ...
\address{$^{4}$Brookhaven National Laboratory, Upton, New York 11973, USA}
% ...
\address{$^{19}$Alikhanov Institute for Theoretical and Experimental Physics, Moscow, Russia}
% ...
\address{$^{21}$Joint Institute for Nuclear Research, Dubna, 141 980, Russia}
\address{$^{22}$Kent State University, Kent, Ohio 44242, USA}
\address{$^{23}$University of Kentucky, Lexington, Kentucky, 40506-0055, USA}
% ...
\address{$^{31}$Ohio State University, Columbus, Ohio 43210, USA}
% ...
\address{$^{34}$Panjab University, Chandigarh 160014, India}
% ...
\address{$^{53}$Variable Energy Cyclotron Centre, Kolkata 700064, India}
% ...



\begin{abstract}
Elliptic flow ($v_{2}$) values for identified particles at mid-rapidity in Au+Au collisions, measured by the STAR experiment in the Beam Energy Scan at RHIC at $\sqrt{s_{NN}}=$ 7.7--62.4 GeV, are presented. A beam-energy dependent difference of the values of $v_{2}$ between particles and corresponding anti-particles was observed. The difference increases with decreasing beam energy and is larger for baryons compared to mesons. This implies that, at lower energies, particles and anti-particles are not consistent with the universal number-of-constituent-quark (NCQ) scaling of $v_{2}$ that was observed at $\sqrt{s_{NN}}=$ 200 GeV.
\end{abstract}


\maketitle
%


Lattice Quantum Chromodynamics (QCD) predicts that at sufficiently high temperatures, $T$,  and/or high baryonic chemical potentials, $\mu_{B}$, normal nuclear matter will undergo a phase transition to a state of matter where quarks and gluons are deconfined, called the Quark-Gluon Plasma (QGP)~\cite{Gross:1980br}.  A Beam Energy Scan (BES) program~\cite{Aggarwal:2010cw} has been carried out at the Relativistic Heavy Ion Collider (RHIC) facility to study the QCD phase structure over a large range in $T$ and $\mu_{B}$.
Particle production in heavy ion collisions with respect to the event plane (EP) can be characterized by the following Fourier expansion:
\begin{equation}
\frac{dN}{d(\phi-\Psi)} \propto 1 + 2 \sum_{n\ge1} v_{n}^{obs} \cos\left[ n(\phi-\Psi) \right],
\label{fPsi_phi}
\end{equation}
where $\phi$ is the azimuthal angle of the particles, $n$ the harmonic number, $v_{n}^{obs}$ the observed Fourier coefficient which has to be corrected for the EP resolution to get $v_{n}$, and $\Psi$ the reconstructed EP azimuthal angle~\cite{Poskanzer:1998yz,Voloshin:1994mz}. The second harmonic coefficient is denoted as elliptic flow, $v_{2}$~\cite{Poskanzer:1998yz}.

Elliptic flow measurements have been used to conclude that strongly interacting partonic matter is produced in Au+Au collisions at $\sqrt{s_{NN}}$ = 200 GeV and that $v_{2}$  develops in the early, partonic, stage. This conclusion is based in part on the observed scaling of $v_{2}$ versus the transverse momentum, $p_{T}$, with the number of constituent-quarks (NCQ)~\cite{Adams:2005zg,Abelev:2007qg,Voloshin:2008dg,Adare:2012vq} for hadrons at intermediate $p_{T}$ (2 to 5 GeV/$c$). Deviations from such a scaling for identified hadron $v_{2}(p_{T})$ at lower beam energies is thus an indication for the absence of a deconfined phase~\cite{Aggarwal:2010cw}.

In a hydrodynamic picture, $v_{2}$ arises in non-central heavy ion collisions due to an initial pressure gradient, which is directly connected to the eccentricity. This leads to particle emission predominantly in the direction of the maximum of the pressure gradient. During the expansion of the system the pressure gradient decreases, which means that elliptic flow primarily probes the early stage of a heavy ion collision. 

For Au+Au collisions at $\sqrt{s_{NN}}$ = 200 GeV, a mass ordering in $v_{2}(p_{T})$ between the different particle species was observed at low transverse momenta ($p_{T}<$ 2 GeV/$c$)~\cite{Adler:2001nb,Adams:2003am,Adams:2005zg}. 
This behaviour can be described by non-viscous hydrodynamic calculations~\cite{Huovinen:2001cy,Nonaka:2003hx,Hirano:2003pw,Shen:2012vn,Schenke,Csanad:2003qa}. The relative mass ordering
can be suppressed by using the reduced transverse kinetic energy ($m_{T}-m_{0}$) instead of $p_{T}$, with $m_{T} = \sqrt{p_{T}^{2} + m_{0}^{2}}$ and $m_{0}$ being the mass of the particle.  At large $(m_{T}-m_{0})$, a splitting in $v_{2}(m_{T}-m_{0})$ 
between baryons and mesons was observed which cannot be described by hydrodynamic calculations. This splitting can be explained, in part, by assuming that the particle production occurs via coalescence of constituent quarks~\cite{Molnar:2003ff}.

The $v_{2}$ values for $\pi^{\pm}$, $K^{\pm}$, $K_{S}^{0}$, $p$, $\bar{p}$, $\phi$, $\Lambda$, $\overline{\Lambda}$, $\Xi^{-}$, $\overline{\Xi}^{+}$, $\Omega^{-}$, $\overline{\Omega}^{+}$ measured at mid-rapidity in minimum bias Au+Au collisions will be reported. The data were recorded by STAR, the Solenoidal Tracker at RHIC, for $\sqrt{s_{NN}}=$ 7.7, 11.5, 19.6, 27, 39, and 62.4 GeV in the years 2010 and 2011 as part of the BES program~\cite{Aggarwal:2010cw}.



\begin{figure*}[]
\centering
\resizebox{17.5cm}{!}{%
  \includegraphics[bb = 0.833555 7.002000 530.879960 166.127971,clip]{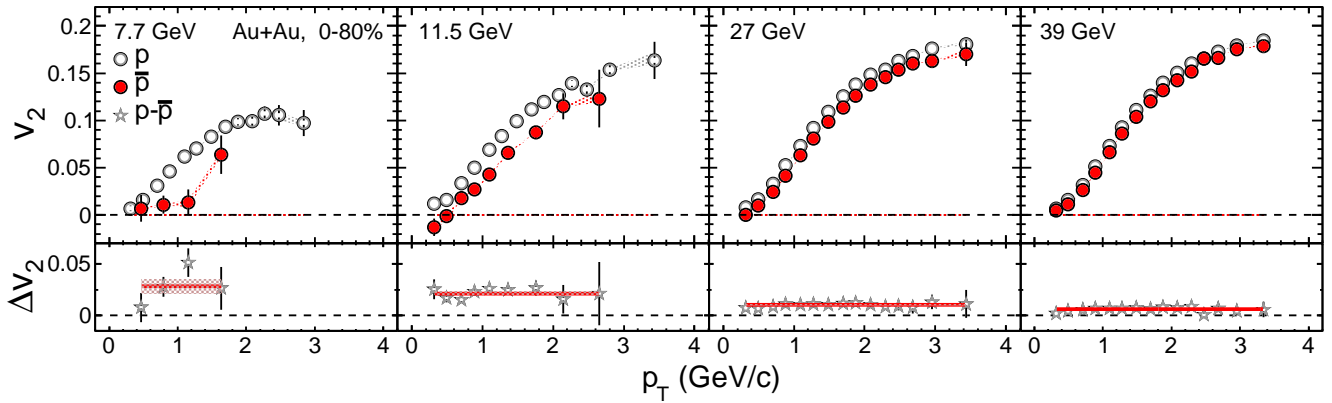}}
  \caption{(Color online) The elliptic flow $v_{2}$ of protons and anti-protons as a function of the transverse momentum, $p_{T}$, for 0--80\% central Au+Au collisions. The lower panels show the difference in $v_{2}(p_{T})$ between the particles and anti-particles. The solid curves are fits with a horizontal line. The shaded areas depict the magnitude of the systematic errors.}
\label{fv2_pt_diff}
\end{figure*}

STAR is a multi-purpose experiment at RHIC with a complete azimuthal coverage. The main detectors used for the data analysis were the Time-Projection Chamber (TPC)~\cite{STAR_TPC} for tracking and particle identification at pseudo-rapidities $|\eta|<1.0$, and the Time-of-Flight (TOF) detector. A minimum bias trigger was defined using a coincidence of hits in the Zero Degree Calorimeters (ZDC), Vertex Position Detectors (VPD), or Beam-Beam Counters (BBC)~\cite{Adamczyk:2012ku}.
To suppress events from collisions with the beam pipe (radius 3.95 cm), an upper limit cut on the radial position of the reconstructed primary vertex of 2 cm was applied. In addition, the z-position of the vertices was limited to values less than $\pm$70 cm. Collisions within a 0--80\% centrality range of the total reaction cross section were selected for the analysis. The centrality definition is based on a comparison between the measured track multiplicity within $|\eta|<0.5$ and a Glauber Monte-Carlo simulation~\cite{Adamczyk:2012ku}.

The particle identification and yield extraction for long-lived charged hadrons ($p$, $\bar{p}$, $\pi^{\pm}$, $K^{\pm}$) was based on a combination of the ionization energy loss, $dE/dx$, in the TPC, the reconstructed momentum ($p$), and the squared mass, $m^{2}$, from the TOF detector~\cite{STAR_BES_PID_v2}. Short-lived particles which decay within the detector acceptance such as $\phi$, $\Lambda$, $\overline{\Lambda}$, $\Xi^{-}$, $\overline{\Xi}^{+}$, $\Omega^{-}$, $\overline{\Omega}^{+}$, and $K_{s}^{0}$ were identified using the invariant mass technique. The combinatorial background to the weakly decaying particles like $\Lambda$ and $\Xi$ was reduced by topological reconstruction. The remaining combinatorial background was fit and subtracted with the mixed event technique~\cite{STAR_BES_PID_v2}.

The event plane was reconstructed using the procedure described in Ref.~\cite{Poskanzer:1998yz}. In order to reduce the effects of non-flow contributions arising mainly from Hanbury-Brown Twiss correlations and Coulomb interactions, the event plane angles were estimated for two sub-events separated by an additional $\eta$-gap instead of using the full TPC event plane method~\cite{STAR_BES_PID_v2}. For such an ``$\eta$-sub-EP" reconstruction, one uses only the particles from the opposite $\eta$ hemisphere with respect to the particle of interest and outside of an additional $\eta$-gap of $|\eta|>0.05$. The non-flow contributions were studied for the six beam energies by comparing different methods of extracting $v_{2}$ for inclusive charged hadrons~\cite{Adamczyk:2012ku}. The four particle cumulant $v_{2}\{4\}$ strongly suppresses non-flow contributions. It has been shown that the difference between $v_{2}(\eta$-${\rm sub})$ and $v_{2}\{4\}$ is 10--20\% for 19.2, 27, and 39 Ge
 V and decreases with decreasing energy. All observed values ($v_{2}^{\rm obs}$) were corrected on an event-by-event basis using the EP resolution~\cite{Masui:2012zh} which was calculated by comparing the two $\eta$-sub-EP angles~\cite{Adamczyk:2012ku}.

For each particle species, the cuts used for particle identification and background suppression were varied to estimate the systematic uncertainties. The errors were also estimated by varying the methods used to flatten the EP, to obtain the yields, and to extract the $v_{2}$ values. A more detailed description of the detector setup and the analysis can be found in Ref.~\cite{STAR_BES_PID_v2}.

\begin{figure}[]
\centering
\resizebox{8cm}{!}{%
\includegraphics[trim=0cm 0cm 0cm 2cm, clip=true, totalheight=0.5\textheight, angle=0]{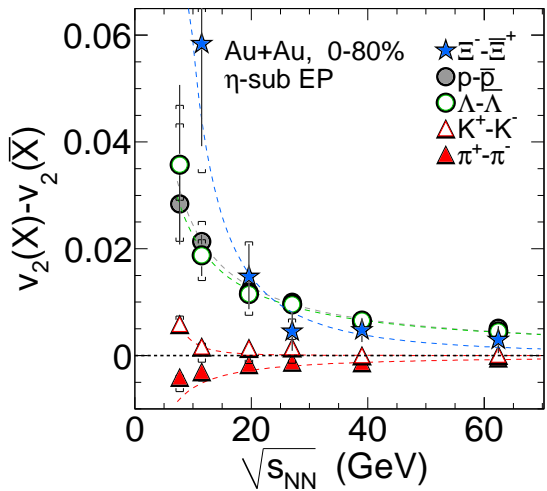}} %
\caption{(Color online) The difference in $v_{2}$ between particles ($X$) and their corresponding anti-particles (${\rm \overline{X}}$) (see legend) as a function of $\sqrt{s_{NN}}$ for 0--80\% central Au+Au collisions. The dashed lines in the plot are fits with a power-law function. The error bars depict the combined statistical and systematic errors.}
\label{fDiff_v2_sNN_muB}
\end{figure}

\begin{figure*}[]
\centering
\resizebox{17.5cm}{!}{%
\includegraphics[bb = 0.000000 25.505999 530.879960 215.711993,clip]{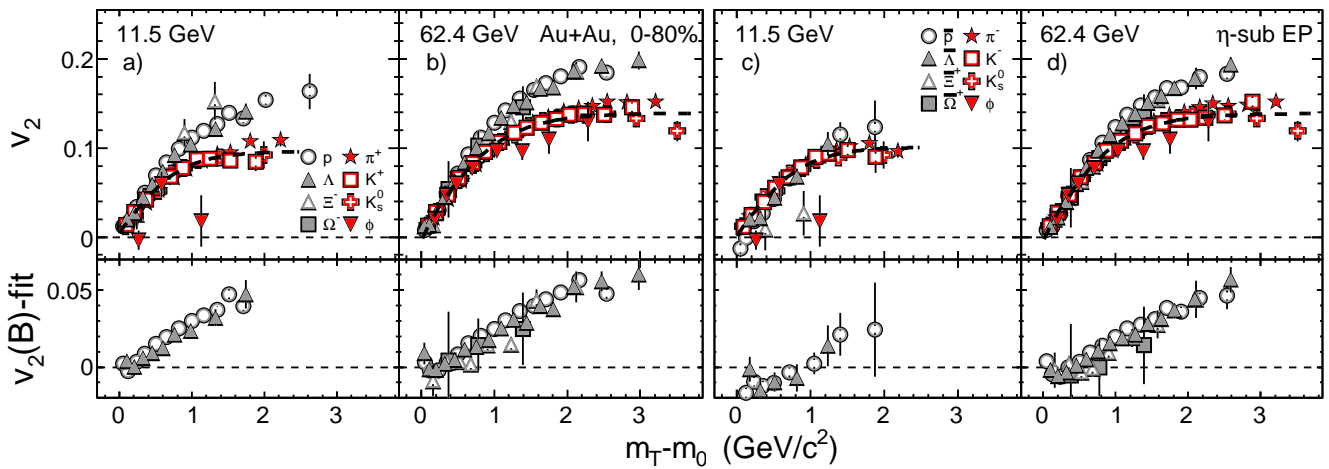}}
\caption{(Color online) The upper panels depict the elliptic flow, $v_{2}$, as a function of reduced transverse mass, $(m_{T}-m_{0})$, for particles, frames a) and b), and anti-particles, frames c) and d), in 0-80\% central Au+Au collisions at $\sqrt{s_{NN}}$ = 11.5 and 62.4 GeV. Simultaneous fits to the mesons except the pions are shown as the dashed lines. The difference of the baryon $v_{2}$ and the meson fits are shown in the lower panels.}
\label{f_v2_mt_diff}
\end{figure*}

In Fig.~\ref{fv2_pt_diff}, the $p_{T}$ dependence of the proton and anti-proton $v_{2}$ is shown for Au+Au collisions at $\sqrt{s_{NN}}$ = 7.7, 11.5, 27, and 39 GeV. At all energies, the $v_{2}$ values increase with increasing $p_{T}$. At $p_{T}=2$ GeV/$c$, the magnitude of  $v_{2}$ for protons increases with energy from about 0.10 at 7.7 GeV to 0.15 at 39 GeV. Lower values of $v_{2}(p_{T})$ are observed for anti-protons compared to protons at all energies. The difference in the $v_{2}$ values for protons and anti-protons increases with decreasing beam energy. The lower panels of Fig.~\ref{fv2_pt_diff} show the $p_{T}$ dependence of the difference in $v_{2}$ for protons and anti-protons. No significant $p_{T}$ dependence is observed, as characterized by the horizontal line fits. The negative values of the anti-proton $v_{2}$ at low $p_{T}$ at $\sqrt{s_{NN}}$ = 11.5 GeV could be due to absorption in the medium \cite{Wang:2012zzi}.

The $v_{2}(p_{T})$ behaviour for $\Lambda (uds)$, $\overline{\Lambda} (\bar{u}\bar{d}\bar{s})$ and $\Xi^{-} (dss)$, $\overline{\Xi}^{+} (\bar{d}\bar{s}\bar{s})$ is similar to that for protons $(uud)$ and anti-protons $(\bar{u}\bar{u}\bar{d})$. In all cases, the baryon anti-particle $v_{2}$ is lower than the corresponding particle $v_{2}$. The $v_{2}(p_{T})$ difference for $\Lambda$ and $\overline{\Lambda}$ is in agreement with previous STAR results at $\sqrt{s_{NN}}$ = 62.4 GeV ~\cite{Abelev:2007qg}. For the mesons $\pi^{+}(u\bar{d})$, $\pi^{-}(\bar{u}d)$, and $K^{+} (u\bar{s})$, $K^{-} (\bar{u}s)$, the differences are smaller than those for the baryons ( the anti-particle convention from~\cite{Beringer:1900zz} is used for mesons). At $\sqrt{s_{NN}}$ = 7.7 GeV, the $v_{2}(p_{T})$ difference between $K^{+}$ and $K^{-}$ is a factor 5--6 smaller as compared to the baryons, with $K^{+}$ having a systematically larger $v_{2}(p_{T})$ than the $K^{-}$. On the other hand, the $v_{2}(p_{T})$ of the $\pi^{-}$ is larger than the $v_{2}(p_{T})$ of the $\pi^{+}$. However, the
  magnitude of the difference for pions as a function of energy is similar to that for the kaons. The details of the $p_{T}$ dependence of the difference in $v_{2}$ between particles and corresponding anti-particles can be found in Ref.~\cite{STAR_BES_PID_v2}.

Figure~\ref{fDiff_v2_sNN_muB} summarizes the variation of the $p_{T}$ independent difference in $v_{2}$ between particles and corresponding anti-particles with $\sqrt{s_{NN}}$. Here, $v_{2}(X)-v_{2}(\overline{X})$ denotes the horizontal line fit values of the difference in $v_{2}(p_{T})$  between particles $X$ ($p$, $\Lambda$, $\Xi^{-}$, $\pi^{+}$, $K^{+}$) and corresponding anti-particles $\overline{X}$ ($\bar{p}$, $\overline{\Lambda}$, $\overline{\Xi}^{+}$, $\pi^{-}$, $K^{-}$). Larger $v_{2}$ values are found for particles than for antiparticles, except for pions for which the opposite ordering is observed. A monotonic increase of the magnitude of $\Delta v_{2} = v_{2}(X)-v_{2}(\overline{X})$ with decreasing beam energy is observed. The data can be described by a power-law function.

While in Au+Au collisions at $\sqrt{s_{NN}}$ = 200 GeV a single NCQ scaling can be observed for particles and anti-particles, the observed difference in $v_2$ at lower beam energies demonstrates that this common NCQ scaling of particles and anti-particles splits. Such a breaking of the NCQ scaling could indicate increased contributions from hadronic interactions in the system evolution with decreasing beam energy.
The energy dependence of $v_{2}(X)-v_{2}(\overline{X})$ could also be accounted for by considering an increase in nuclear stopping power with decreasing $\sqrt{s_{NN}}$ if the $v_{2}$ of transported quarks (quarks coming from the incident nucleons) is larger than the $v_{2}$ of produced quarks~\cite{Dunlop:2011cf,Steinheimer:2012bn}.
Theoretical calculations~\cite{Xu:2012gf} suggest that the difference between particles and anti-particles could be accounted for by mean field potentials where the $K^{-}$ and $\bar{p}$ feel an
attractive force while the $K^{+}$ and $p$ feel a repulsive force. 
Most of the current theoretical calculations can reproduce the basic pattern, but fail to quantitatively reproduce the measured $v_2$ difference \cite{Dunlop:2011cf,Steinheimer:2012bn,Xu:2012gf,Song:2012cd}. So far, none of the theory calculations describes the observed ordering of the particles. Therefore, more accurate calculations from theory are needed to distinguish between the different possibilities. Other possible interpretations for the observation that the $\pi^{-}$ $v_{2}(p_{T})$ is larger than the $\pi^{+}$ $v_{2}(p_{T})$ is the Coulomb repulsion of $\pi^{+}$ by the mid-rapidity net-protons (only at low $p_{T}$) and the chiral magnetic effect in finite baryon-density matter~\cite{Burnier:2011bf}.
In Ref.~\cite{STAR_BES_PID_v2}, the study of the centrality dependence of $\Delta v_{2}$ for protons and anti-protons is extended to investigate, if different production rates for protons and anti-protons as a function of centrality could cause the observed differences. It was observed that the differences, $\Delta v_{2}$, are significant at all centralities.

\begin{figure*}[]
\centering
\resizebox{17.5cm}{!}{%
  \includegraphics[bb = 0.000000 25.505999 530.879960 215.711993,clip]{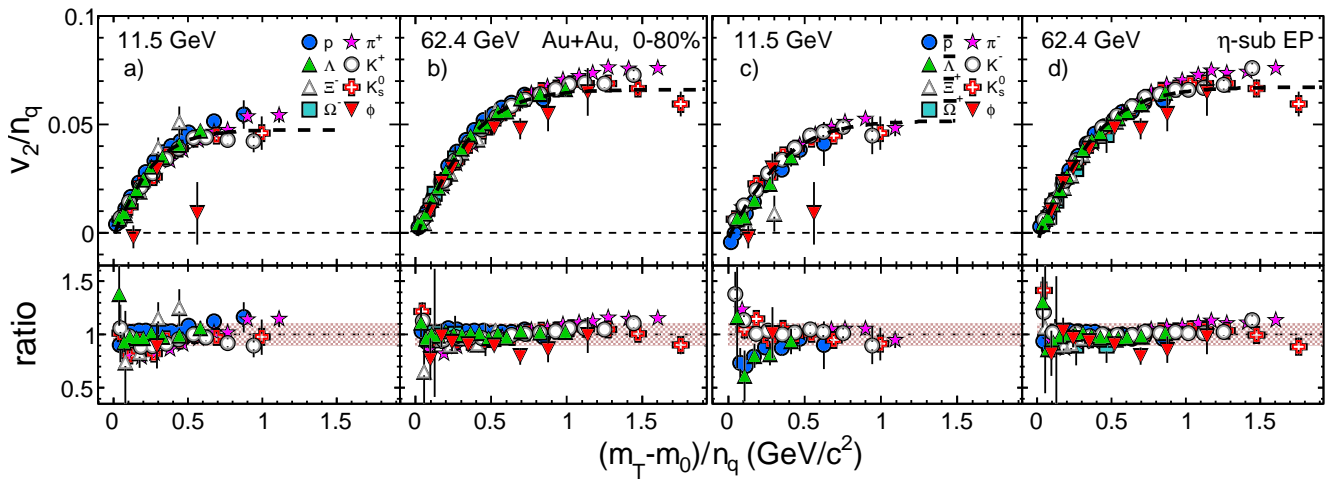}}
  \caption{(Color online) The number-of-constituent quark scaled elliptic flow $(v_{2}/n_{q})((m_{T}-m_{0})/n_{q})$ for 0--80\% central Au+Au collisions at $\sqrt{s_{NN}}$ = 11.5 and 62.4 GeV for selected particles, frames a) and b), and corresponding anti-particles, frames c) and d). The dashed lines are simultaneous fits~\cite{Dong:2004ve} to all of the data sets at a given energy. The lower panels depict the ratios to the fits, while a $\pm$10\% interval is shown as the shaded area to guide the eye. Some data points for $\phi$ and $\Xi$ are out of the plot range in the lower panels of frames a) and c).}
\label{fNCQ_scale}
\end{figure*}

The $v_{2}(m_{T}-m_{0})$ and possible NCQ scaling was also investigated for particles and anti-particles separately.  
Figure~\ref{f_v2_mt_diff} shows $v_{2}$ as a function of the reduced transverse mass, $(m_{T}-m_{0})$, for various particles and anti-particles at $\sqrt{s_{NN}}$ = 11.5 and 62.4 GeV. 
The baryons and mesons are clearly separated for $\sqrt{s_{NN}}$ = 62.4 GeV at $(m_{T}-m_{0}) > 1$  GeV/$c^{2}$. 
While the effect is present for particles at $\sqrt{s_{NN}}$ = 11.5 GeV, no such separation is observed for the anti-particles at this energy in the measured $(m_{t}-m_{0})$ range up to 2 GeV/$c^{2}$. 
The lower panels of Fig.~\ref{f_v2_mt_diff} depict the difference of the baryon $v_{2}$ relative to a fit to the meson $v_{2}$ data with the pions excluded from the fit. 
The anti-particles at $\sqrt{s_{NN}}$ = 11.5 GeV show a smaller difference compared to the particles.

In Fig.~\ref{fNCQ_scale}, the $v_{2}(m_{T}-m_{0})$ values scaled on both axes with the number of constituent-quarks are presented for $\sqrt{s_{NN}}$ = 11.5 and 62.4 GeV. A simultaneous fit~\cite{Dong:2004ve} to all data sets at a given energy
is shown as the dashed black line. The ratio of the data to these fits is shown in the lower panels. Most of the ratio values  
are within $\pm$10\% of unity, showing that the NCQ scaling holds for the group of particles while for the group of anti-particles the measured kinematic range is not enough to conclude. The $\phi$ mesons are an exception.  At the highest $(m_{T}-m_{0})/n_{q}$ values, the $\phi$ meson data point for $\sqrt{s_{NN}}$ = 11.5 GeV ($p_{T} = 1.9$ GeV/$c$) is  2.3$\sigma$ lower than those of the other hadrons. This is comparable to the observed deviation at $\sqrt{s_{NN}}$ = 7.7 GeV ($p_{T} = 1.7$ GeV/$c$) by 1.8$\sigma$~\cite{STAR_BES_PID_v2}. The smaller $v_{2}$ values of the $\phi (s\bar{s})$ meson, which has a smaller hadronic interaction cross section~\cite{Sibirtsev:2006yk}, may indicate that hadronic interactions become more important than partonic effects for the systems formed at collision energies $\lesssim$ 11.5 GeV~\cite{Mohanty:2009tz,Nasim:2013fb}. 

In summary, the first observation of a beam-energy dependent difference in $v_{2}(p_{T})$ between particles 
and corresponding anti-particles for minimum bias $\sqrt{s_{NN}}$ = 7.7--62.4 GeV Au+Au collisions at mid-rapidity is reported.  
The difference increases with decreasing beam energy. Baryons show a larger difference compared to mesons. The relative values of $v_{2}$ for charged pions have the opposite trend to the values of charged kaons.  
It is concluded that, at the lower energies, particles and anti-particles are no longer consistent with the single NCQ scaling that was observed for $\sqrt{s_{NN}}$ = 200 GeV. However, for the group of particles the NCQ scaling holds within $\pm$10\% while for the group of anti-particles the difference between baryon and meson $v_{2}$ continues to decrease to lower energies. We further observed that the $\phi$ meson $v_2$ at the highest measured  $m_{T}-m_{0}$ value is low compared to other hadrons at $\sqrt{s_{NN}}$ = 7.7 and 11.5 GeV with 1.8$\sigma$ and 2.3$\sigma$, respectively. 

We thank the RHIC Operations Group and RCF at BNL, the NERSC Center at LBNL and the Open Science Grid consortium for providing resources and support. This work was supported in part by the Offices of NP and HEP within the U.S. DOE Office of Science, the U.S. NSF, the Sloan Foundation, CNRS/IN2P3, FAPESP CNPq of Brazil, Ministry of Ed. and Sci. of the Russian Federation, NNSFC, CAS, MoST, and MoE of China, GA and MSMT of the Czech Republic, FOM and NWO of the Netherlands, DAE, DST, and CSIR of India, Polish Ministry of Sci. and Higher Ed., Korea Research Foundation, Ministry of Sci., Ed. and Sports of the Rep. of Croatia, and RosAtom of Russia, and VEGA of Slovakia.



\end{document}